\documentclass[twocolumn, superscriptaddress, showpacs, pra]{revtex4-1}

\usepackage{color,amsbsy,amssymb,latexsym,amsfonts, amsmath,cancel}
\usepackage{braket}
\usepackage{mathrsfs}
\usepackage{graphicx}

\newcommand{\bel}[1]{\begin{equation}\label{#1}}
\newcommand{\bal}[1]{\begin{eqnarray}\label{#1}}
\newcommand{\be}{\begin{equation}}
\newcommand{\ba}{\begin{eqnarray}}
\newcommand{\ee}{\end{equation}}
\newcommand{\ea}{\end{eqnarray}}

\newcommand{\bea}{\begin{equation}}
\newcommand{\eea}{\end{equation}}

\begin{document}

\date{June 20, 2013}

\title{Weak value expansion of quantum operators and its application in stochastic matrices}

\author{Taksu Cheon}
\email[Email:]{taksu.cheon@kochi-tech.ac.jp}
\homepage{http://researchmap.jp/T\!\_Zen/}
\affiliation{Laboratory of Physics, Kochi University of Technology, Tosa Yamada, Kochi 782-8502, Japan}
\author{Sergey Poghosyan}
\email[Email:]{sergey.poghosyan@kochi-tech.ac.jp}
\homepage{http://researchmap.jp/Sergey.Poghosyan/}
\affiliation{Laboratory of Physics, Kochi University of Technology, Tosa Yamada, Kochi 782-8502, Japan}

\begin{abstract}
It is shown that any Hermitian operator can be expanded in terms of a set of operators formed from biorthogonal basis, and the expansion coefficients are given as products of weight functions and weak values, shedding a new light on the physical interpretation of the weak value.  The utility of our approach is showcased with examples of spin one-half and spin one systems, where irreversible subset of stochastic matrices describing projective measurement on a mixed state is identified.
\end{abstract}

\maketitle

\section{Introduction}

The concept of weak value, along with its experimental validation by weak measurements,
has been around for almost quarter century \cite{AA88,AA90,RHS91,RA95,AA2005}.
It is based on the idea of physical value subjected to two successive measurements \cite{Aharonov_1964},
which has been later extended to general multiple-time measurements \cite{AP09}.
The weak value formalism has a wide range of applications in various fields
of quantum information theory. Particularly, it can be used to transfer quantum-communication protocols \cite{BR2000},
describe entangled systems \cite{ABPR2002}, reconstruct quantum optical states
by weak measurements \cite{HLS11,FF12} and clone quantum systems \cite{Hof12}.
To many physicists' minds, however, the concept still entails an aura of mystery, and confusion over its physical interpretations \cite{Ta02, SH10, MS12} never seems to have been fully cleared.

In this article, we intend to reconcile this often mystified concept of weak value with the conventional orthogonal vector space formulation of quantum mechanics by introducing a {\em complete set} of weak values.
We point out the existence of a set of operators defined by biorthogonal Hilbert vector bases, with which any Hermitian operator can be expanded, upon which the weak values emerge as the expansion coefficients.
One particular weak value can dominate over all others when certain conditions are met, presenting the phenomenon of ``magnification of physical value'' with weak measurement.
We illustrate our argument with the examples of spin one-half and spin one systems.

Virtue of bringing in the whole set of weak values, as opposed to one particular weak value,
is demonstrated by another example, again involving spin one-half and spin one systems,
in which a mixed quantum state undergoes a projective measurement that brings the system into another mixed state.
This process is shown to be  easily described by an unistochastic matrix, a ``quantum'' subset of stochastic
matrices \cite{Bi46}.  We identify the condition in which the unistochastic matrix is irreversible for the
case of a spin one-half and spin one systems.  We obtain the subset of all possible projective measurements, for
which the history is erased, namely, the reconstruction of the original state is impossible after the measurement.

\section{Weak value expansion}
Consider a Hermitian operator ${ A}$ on Hilbert space of dimension $n$.
We attempt to represent ${ A}$ with two different orthonormal bases
%
$\{\left| \psi_j \right>, j=1, ..., n \}$ with
$\left< \psi_i | \psi_j \right>=\delta_{i,j}$  and
$\{\left| \phi_\ell\right>, \ell=1, ..., n \}$ with
$\left< \phi_k | \phi_\ell \right>=\delta_{k,\ell}$.
%
We assume that we have the property $\left< \phi_\ell | \psi_j \right> \ne 0 \ {\rm for\ all } \ \ell, j$.
Following Aharonov, Albert and Vaidman \cite{AA88}, let us define the weak value $(A)$ of ${ A}$  by
\begin{eqnarray}
(A)_{\ell,j} = \frac{\left< \phi_\ell \right| { A} \left|  \psi_j \right> }{\left< \phi_\ell | \psi_j \right>} .
\end{eqnarray}
We also define, in the manner of Reznik and Aharonov \cite{RA95}, the $W$ operator by
\begin{eqnarray}
{ W}_{\ell,j} = \frac{\left|  \phi_\ell \right> \left< \psi_j  \right|}{\left< \psi_j  | \phi_\ell \right>} ,
\end{eqnarray}
which is a biorthogonal basis extension of the density matrix operator.
We also define the overlap matrices
\begin{eqnarray}
{\mu}_{\ell,j} = \left| {\left< \phi_\ell | \psi_j \right>} \right| ^2 .
\end{eqnarray}
We have, with straightforward calculation, a unique expansion of the operator ${ A}$ in terms of $n^2$ set of operators  ${W}_{\ell,j}$ ;
\begin{eqnarray}
\label{wkexp}
{ A} = \sum_{\ell,j} (A)_{\ell,j} \, { W}_{\ell,j}  \, {\mu}_{\ell,j} .
\end{eqnarray}
Note the relation
\begin{eqnarray}
{ W}_{\ell,j} { W}^\dagger_{j', \ell'}
= \delta_{j',j}  \frac{ \left|  \phi_{\ell'} \right>  \left< \phi_\ell \right|}
   { \left< \phi_\ell | \psi_j \right> \left<  \psi_{j'} | \phi_{\ell'} \right> } ,
\end{eqnarray}
which leads to the orthogonality relation
\begin{eqnarray}
 \left< \phi_k \right| { { W}_{\ell,j} { W}^\dagger_{j', \ell'}  } \left|  \phi_k  \right>
= \delta_{j',j} \delta_{\ell,k} \delta_{\ell',k}
   \frac{1}{ \mu_{k, j}} .
\end{eqnarray}
From this, we easily have the formula to obtain the weak value of ${ A}$ with ${ W}$ operators using the trace;
\begin{eqnarray}
\label{wexp7}
(A)_{\ell,j}  = {\rm tr} [{ A} { W}^\dagger_{j, \ell} ]
 = {\rm tr} [{ W}_{\ell, j} { A}  ] .
\end{eqnarray}
We conclude that the set of $n^2$ weak values $(A)_{\ell,j}$ $(\ell, j = 1, .., n)$ characterizes the operator ${ A}$ completely.

We can also consider two mixed states
\begin{eqnarray}
{ \rho}_{p} = \sum_{ j}   {\left| \psi_j \right> p_j \left< \psi_j \right|} ,
\quad
{ \rho}_{q} = \sum_{ \ell}   {\left| \phi_\ell \right> q_j \left< \phi_\ell \right|} ,
\end{eqnarray}
for which,  the $W$ operator is given by
\begin{eqnarray}
{ W}_{q,p} = \sum_{\ell, j} q_\ell p_j \frac{\left| \phi_\ell \right> \left< \psi_j \right|}{\left< \psi_j | \phi_\ell \right>} .
\end{eqnarray}
We obtain the mixed state weak value of a Hermitian operator $A$ from this $W$ operator as
\begin{eqnarray}
(A)_{q,p}  = {\rm tr} [ { W}_{q,p} { A} ] = \sum_{\ell,j} q_\ell p_j   (A)_{\ell,j} .
\end{eqnarray}
This relation shows how the relative phases of weak values of different post and preselection states are to be experimentally determined.
%

%
The weak value expansion is, technically,  just a simple modification of standard expansion of a Hermitian operator ${ A}$ with orthonormal states $\{ \left| \psi_j \right> \}$
\begin{eqnarray}
{ A} = \sum_{i, j} \left| \psi_i \right> \left< \psi_i \right| { A}\left| \psi_j \right> \left< \psi_j \right| ,
\end{eqnarray}
using two sets of orthogonal states $\{ \left| \psi_j \right>,  \left| \phi_\ell \right>  \}$ instead of one set.
As such, the validity of weak values as a set of observable quantities which all together form a full description of the operator ${ A}$, eq. (\ref{wkexp}), is quite independent from various ``interpretations'' of the weak value, that has ranged from practical to metaphysical.
An important property of the weak value expansion is that all quantities appearing in (\ref{wkexp}) are ``gauge invariant'' in the sense of invariance with respect to the change of phases of orthonormal vectors, $\left| \psi_j \right> \to e^{i \chi_j}\left| \psi_j \right>$ and $\left| \phi_\ell \right> \to e^{i \xi_j}\left| \phi_\ell \right>$.  This contrasts with the traditional treatment of non diagonal matrix elements of Hermitian operator $\left< \psi_j | A | \psi_k \right>$ whose gauge dependence renders them unsuitable for direct measurements, and the transition probabilities are related only with their absolute squares.

The special feature of the weak value expansion is the fact that the weak value $(A)_{\ell,j}$ takes the eigenvalue
of $\left| \psi_j \right>$ when ${ A}$ is diagonal in the basis $\{ \left| \psi_j \right> \}$, and the eigenvalue of
$\left| \phi_\ell \right>$ when ${ A}$ is diagonal in the basis $\{  \left| \phi_\ell \right> \}$.
When $\{  \left| \psi_j \right> \}$ and $\{  \left| \phi_\ell \right> \}$ are identified as the states of a system
at certain two times $t_1$ and $t_2$, and if ${ A}$ is some ``moderate mixture'' of two operators, which are respectively
diagonal in  $\{  \left| \psi_j \right> \}$ and $\{  \left| \phi_\ell \right> \}$, it may therefore be possible to
interpret the weak value  $(A)_{\ell,j}$ as the "physical value" of the operator ${ A}$ at any time in between $t_1$ and $t_2$.
But of course, the validity of such interpretation is quickly lost when such condition is not met.

A useful expression
\begin{eqnarray}
\left< \psi_k \right| { A} \left| \psi_k \right> =  \sum_{\ell} (A)_{\ell,k} \  \, {\mu}_{\ell,k}
\end{eqnarray}
is obtained from (\ref{wkexp}) straightforwardly.  Since ${\mu}_{\ell,k}$ is the probability of finding the state $\left| \phi_\ell \right>$ in the state $\left| \psi_k \right>$, the weak value $(A)_{\ell,k}$ is interpretable as $\ell$-th factional component of the expectation value $\left< \psi_k \right| { A} \left| \psi_k \right>$ in terms of post-selection states.
Similarly, we also obtain
\begin{eqnarray}
\left< \phi_m \right| { A} \left| \phi_m \right> =  \sum_{j} (A)_{m, j} \  \, {\mu}_{m, j} ,
\end{eqnarray}
with analogous interpretation as  $j$-th factional component of the expectation value $\left< \psi_k \right| { A} \left| \psi_k \right>$ in terms of pre-selection states.

\section{Examples with spin matrices}
\subsection{Spin one-half}
\subsubsection{Maximally exclusive choice for pre- and post-selection states}
An example of Pauli spin operators  with a set $\{$ $\left| \psi_1 \right> = \left| \uparrow \right>$, $\left| \psi_2 \right> = \left| \downarrow \right>$ $\}$ and $\{$ $\left| \phi_1 \right> = \left| \rightarrow \right>$, $\left| \phi_2 \right> = \left| \leftarrow \right>$ $\}$, namely
\begin{eqnarray}
&
\left| \psi_1 \right> = \begin{pmatrix} 1 \\ 0 \end{pmatrix} ,
\left| \psi_2 \right> = \begin{pmatrix} 0 \\ 1 \end{pmatrix} ,
\nonumber \\ &{\rm and} \nonumber \\
&
\left| \phi_1 \right> = \frac{1}{\sqrt{2}}\begin{pmatrix} 1 \\ 1 \end{pmatrix} ,
\left| \phi_2 \right> = \frac{1}{\sqrt{2}}\begin{pmatrix} 1 \\ -1 \end{pmatrix} ,
\end{eqnarray}
 should be instructive.  We have, for all $\ell$, $j$,
\begin{eqnarray}
\mu_{\ell,j} = \frac{1}{2}.
\end{eqnarray}
An explicit matrix representation of ${ W}$ operators are
\begin{eqnarray}
&&\!\!\!\!\!\!\!\!\!\!\!\!
{ W}_{1,1} = \begin{pmatrix} 1 & 0 \\ 1 & 0 \end{pmatrix} ,
\quad
{ W}_{1,2} = \begin{pmatrix} 0 & 1 \\ 0 & 1 \end{pmatrix},
\nonumber \\
&&\!\!\!\!\!\!\!\!\!\!\!\!
{ W}_{2,1} = \begin{pmatrix} 1 & 0 \\ -1 & 0 \end{pmatrix} ,
\quad
{ W}_{2,2} =  \begin{pmatrix} 0 & -1 \\ 0 & 1 \end{pmatrix} .
\end{eqnarray}
The weak values of $\sigma_x$ and $\sigma_z$ are given by
\begin{eqnarray}
&
(\sigma_x)_{1,1} = 1, \  (\sigma_x)_{1,2} = 1,
\nonumber \\
&
(\sigma_x)_{2,1} = -1, \  (\sigma_x)_{2,2} = -1,
\end{eqnarray}
and
\begin{eqnarray}
&
(\sigma_z)_{1,1} = 1, \quad (\sigma_z)_{1,2} = -1,
\nonumber \\
& (\sigma_z)_{2,1} = 1, \quad (\sigma_z)_{2,2} = -1 ,
\end{eqnarray}
which is easily understood in terms of post-selection $(A)_{\ell, \bullet}$ and pre-selection $(A)_{\bullet, j}$.
If we consider the ``intermediate'' spin operator
\begin{eqnarray}
\sigma_\theta = \sigma_z \cos{\theta} +\sigma_x \sin{\theta} ,
\end{eqnarray}
its weak values are given by
\begin{eqnarray}
&&\!\!\!\!\!\!\!\!\!\!\!\!
(\sigma_\theta)_{1,1} = \cos{\theta} + \sin{\theta}, \quad (\sigma_\theta)_{1,2} = \cos{\theta} - \sin{\theta},
\nonumber \\
&&\!\!\!\!\!\!\!\!\!\!\!\!
(\sigma_\theta)_{2,1} = -\cos{\theta} + \sin{\theta}, \quad (\sigma_\theta)_{2,2} = -\cos{\theta} - \sin{\theta},
\end{eqnarray}
which smoothly interpolates $(\sigma_z)$ and $(\sigma_x)$.  Still, they can attain the maximum value $\sqrt{2}$ going beyond the classically allowed value of 1.
The weak values of $\sigma_y$ are
\begin{eqnarray}
&
(\sigma_y)_{1,1} = {\rm i}, \quad (\sigma_y)_{1,2} = -{\rm i},
\nonumber \\
&(\sigma_y)_{2,1} = -{\rm i}, \quad (\sigma_y)_{2,2} = {\rm i},
\end{eqnarray}
whose imaginariness and counter intuitive sign assignments are the signature of  quantum incompatibility of the measurement of $\sigma_y$ with both post- and pre-selection basis.
The expansions of $\sigma$ matrices in terms of $W$ operators are given by
\begin{eqnarray}
&
\sigma_x = \frac{1}{2}
 \left( { W}_{1,1} + { W}_{1,2} -{ W}_{2,1} -{ W}_{2,2}  \right) ,
\nonumber \\
&
\sigma_y = \frac{{\rm i}}{2}
 \left( { W}_{1,1} - { W}_{1,2} -{ W}_{2,1} +{ W}_{2,2}  \right) ,
\nonumber \\
&
\sigma_z = \frac{1}{2}
 \left( { W}_{1,1} - { W}_{1,2} +{ W}_{2,1} -{ W}_{2,2}  \right) ,
\end{eqnarray}
which are, in a sense, a trivial, but an instructive relations.

\subsubsection{Generic choice for pre- and post- selection states}
Alternative choice for the post-selection states
\begin{eqnarray}
\left| \phi_1 \right> = \begin{pmatrix} \cos\frac{\theta}{2} \\ \sin\frac{\theta}{2} \end{pmatrix} ,
\left| \phi_2 \right> = \begin{pmatrix} -\sin\frac{\theta}{2} \\ \cos\frac{\theta}{2} \end{pmatrix} ,
\end{eqnarray}
with the same pre-selection states as before
\begin{eqnarray}
\left| \psi_1 \right> = \begin{pmatrix} 1 \\ 0 \end{pmatrix} ,
\left| \psi_2 \right> = \begin{pmatrix} 0 \\ 1 \end{pmatrix} ,
\end{eqnarray}
is also revealing.  In this case, for the weight matrix, we have 
\begin{eqnarray}
\mu_{1,1} = \mu_{2,2} = \cos^2 \frac{\theta}{2}, \quad
\mu_{1,2} = \mu_{2,1} = \sin^2 \frac{\theta}{2}.
\end{eqnarray}
The $W$-matrix is given by
\begin{eqnarray}
&
{ W}_{1,1} = \begin{pmatrix}  1 & 0  \\ \tan\frac{\theta}{2} & 0 \end{pmatrix} ,
\quad
{ W}_{1,2} = \begin{pmatrix} 0 & \cot\frac{\theta}{2} \\ 0 & 1  \end{pmatrix},
\nonumber \\
&
{ W}_{2,1} = \begin{pmatrix}  1 & 0  \\ -\cot\frac{\theta}{2} & 0 \end{pmatrix} ,
\quad
{ W}_{2,2} =  \begin{pmatrix}  0 & -\tan\frac{\theta}{2} \\ 0  & 1 \end{pmatrix} .
\end{eqnarray}
Note that some of the elements diverge at $\theta \to 0$.  They are to be  compensated by the weight function to give finite answer to all physical quantities.
The weak values of Pauli matrix $\sigma_z$ are given by
\begin{eqnarray}
&
(\sigma_z)_{1,1} = 1, \quad (\sigma_z)_{1,2} = -1,
\nonumber \\
&
(\sigma_z)_{2,1} = 1, \quad (\sigma_z)_{2,2} = -1,
\end{eqnarray}
which are easily understood because the pre-selection states $\psi_{j}$ are the eigenstates of this operator.
For $\sigma_x$, we have the weak values
\begin{eqnarray}
&
(\sigma_x)_{1,1} = \tan\frac{\theta}{2}, \quad (\sigma_x)_{1,2} = \cot\frac{\theta}{2},
\nonumber \\
&
(\sigma_x)_{2,1} = -\cot\frac{\theta}{2}, \quad (\sigma_x)_{2,2} = -\tan\frac{\theta}{2} ,
\end{eqnarray}
some of which diverge at $\theta \to 0$, indicating the existence of the weak measurement amplification.  The meaning of this amplification is understood by calculating the weak values of rotated spin operator
\begin{eqnarray}
\sigma_\theta = \sigma_z \cos{\theta} +\sigma_x \sin{\theta},
\end{eqnarray}
which is given by
\begin{eqnarray}
&
(\sigma_\theta)_{1,1} = 1, \quad (\sigma_\theta)_{1,2} = 1,
\nonumber \\
&
(\sigma_\theta)_{2,1} = -1, \quad (\sigma_\theta)_{2,2} = -1,
\end{eqnarray}
which simply reflects the fact that the post-selection states $\phi_{\ell}$ are the eigenstates of $\sigma_\theta$ operator.   The relation
\begin{eqnarray}
\sigma_x = - \sigma_z \cot{\theta} +\sigma_\theta {\rm \,cosec\,}{\theta}
\end{eqnarray}
tells the source of divergence of $(\sigma_x)$ with this choice of post selection states at $\theta \to 0$.
The weak values of $\sigma_y$ are
\begin{eqnarray}
&
(\sigma_y)_{1,1} = {\rm i}\tan\frac{\theta}{2}, \quad (\sigma_y)_{1,2} = -{\rm i}\cot\frac{\theta}{2},
\nonumber \\ &
(\sigma_y)_{2,1} = -{\rm i}\cot\frac{\theta}{2}, \quad (\sigma_y)_{2,2} = {\rm i} \tan\frac{\theta}{2}.
\end{eqnarray}
The expansions of $\sigma$ matrices in terms of $W$ operators now are given by
\begin{eqnarray}
&
\sigma_z =
  { W}_{1,1} \cos^2 \frac{\theta}{2}
- { W}_{1,2} \sin^2 \frac{\theta}{2}
  \qquad \qquad\qquad\qquad \nonumber \\ &
\qquad\qquad\qquad
 + { W}_{2,1} \sin^2 \frac{\theta}{2}
-{ W}_{2,2} \cos^2 \frac{\theta}{2} ,
\nonumber \\
&
\sigma_y =
  {\rm i} \tan\frac{\theta}{2} { W}_{1,1} \cos^2 \frac{\theta}{2}
-{\rm i} \cot\frac{\theta}{2} { W}_{1,2} \sin^2 \frac{\theta}{2}
 \qquad\quad \nonumber \\ &
\
-{\rm i} \cot\frac{\theta}{2} { W}_{2,1} \sin^2 \frac{\theta}{2}
+{\rm i} \tan\frac{\theta}{2}{ W}_{2,2} \cos^2 \frac{\theta}{2} ,
\nonumber \\
&
\sigma_x =
  \tan\frac{\theta}{2} { W}_{1,1} \cos^2 \frac{\theta}{2}
+ \cot\frac{\theta}{2} { W}_{1,2} \sin^2 \frac{\theta}{2}
 \qquad\qquad \nonumber \\ &
\quad\
-\cot\frac{\theta}{2} { W}_{2,1} \sin^2 \frac{\theta}{2}
-  \tan\frac{\theta}{2}{ W}_{2,2} \cos^2 \frac{\theta}{2} ,
\end{eqnarray}
which clearly show the nature of weak value expansion.
%
\subsection{Spin one}
We now take a look at the example of a three-state Hilbert space.
Consider the operators $L_x$, $L_y$, $L_z$ defined by the algebra
\begin{eqnarray}
L_i L_j - L_j L_i = i \varepsilon_{i,j,k} L_k.
\end{eqnarray}
Let us take an explicit representation
\begin{eqnarray}
&
L_x = \frac{1}{\sqrt{2}}\begin{pmatrix} 0 & 1 & 0 \\ 1 & 0 & 1 \\ 0 & 1 & 0 \end{pmatrix},
L_y = \frac{{\rm i}}{\sqrt{2}}\begin{pmatrix} 0 & -1 & 0 \\ 1 & 0 & -1 \\ 0 & 1 & 0 \end{pmatrix},
\nonumber \\ &
L_z = \begin{pmatrix} 1 & 0 & 0 \\ 0 & 0 & 0 \\ 0 & 0 & -1 \end{pmatrix}.
\end{eqnarray}
We take the pre-selection states as
\begin{eqnarray}
\left| \psi_1 \right> = \begin{pmatrix} 1 \\ 0 \\ 0 \end{pmatrix} ,
\left| \psi_2 \right> = \begin{pmatrix} 0 \\ 1 \\ 0 \end{pmatrix} ,
\left| \psi_3 \right> = \begin{pmatrix} 0 \\ 0 \\ 1 \end{pmatrix} ,
\end{eqnarray}
and the post-selection states as
\begin{eqnarray}
\left| \phi_1 \right> =
\begin{pmatrix} \cos^2\frac{\theta}{2} \\  \frac{\sin\theta}{\sqrt{2}} \\ \sin^2\frac{\theta}{2} \end{pmatrix} ,
\left| \phi_2 \right> =
\begin{pmatrix} -\ \frac{\sin\theta}{\sqrt{2}} \\  \cos\theta \\  \frac{\sin\theta}{\sqrt{2}} \end{pmatrix} ,
\left| \phi_3 \right> =
\begin{pmatrix} \sin^2\frac{\theta}{2} \\  -\frac{\sin\theta}{\sqrt{2}} \\ \cos^2\frac{\theta}{2} \end{pmatrix} ,
\nonumber \\
\end{eqnarray}
which are obtained from $\left| \psi_j \right>$ by rotation around $y$-axis by angle $\theta$.
Namely, they are the eigenstates of the operator $L_\theta$ defined
\begin{eqnarray}
L_\theta = L_z \cos{\theta} +L_x \sin{\theta} ,
\end{eqnarray}
with eigenvalues $+1$, $0$, and $-1$, respectively.  In an analogous manner to the previous example of spin one-half, we can consider the complete set of weak values of arbitrary operator on the Hilbert space using the pre-selection states $\left| \psi_j \right>$ and post-selection states $\left| \phi_\ell \right>$.
We first calculate the weight matrix,
which is given by
\begin{eqnarray}
\mu_{1,1} = \cos^4 \frac{\theta}{2}, \
\mu_{1,2} = \frac{\sin^2 \theta}{2}, \
\mu_{1,3} = \sin^4 \frac{\theta}{2},
\nonumber \\
\mu_{2,1} = \frac{\sin^2 \theta}{2}, \
\mu_{2,2} = \cos^2 {\theta}, \
\mu_{2,3} =\frac{\sin^2 \theta}{2},
\nonumber \\
\mu_{3,1} = \sin^4 \frac{\theta}{2}, \
\mu_{3,2} = \frac{\sin^2 \theta}{2}, \
\mu_{3,3} = \cos^4 \frac{\theta}{2}.
\end{eqnarray}
The $W$-matrix is given by
\begin{widetext}
\begin{eqnarray}
&
{ W}_{1,1} = \begin{pmatrix}   1 & 0 &0  \\ \sqrt{2} \tan\frac{\theta}{2} & 0 & 0 \\ \tan^2\frac{\theta}{2} & 0 & 0\end{pmatrix} ,
{ W}_{2,1} = \begin{pmatrix}  0 & \frac{\cot\frac{\theta}{2}}{\sqrt{2}} & 0 \\0 & 1 & 0  \\
   0 & \frac{\tan\frac{\theta}{2}}{\sqrt{2}} & 0\end{pmatrix} ,
{ W}_{1,3} = \begin{pmatrix}   0 & 0 &  \cot^2\frac{\theta}{2} \\ 0 & 0 & \sqrt{2} \cot\frac{\theta}{2} \\
 0 &  0 & 1  \end{pmatrix} ,
\nonumber \\
&
{ W}_{2,1} = \begin{pmatrix}  1 & 0 &  0 \\ - \sqrt{2} \cot\theta & 0 & 0 \\ -1 & 0 & 0\end{pmatrix} ,
{ W}_{2,2} = \begin{pmatrix} 0 & - \frac{\tan\theta}{\sqrt{2}} & 0\\ 0 &  1 & 0  \\ 0 & \frac{\tan\theta}{\sqrt{2}} & 0 \end{pmatrix} ,
{ W}_{2,3} = \begin{pmatrix} 0 & 0 & -1 \\ 0 & 0 & \sqrt{2} \cot\theta\\ 0 &  0 &  1 \end{pmatrix} ,
\nonumber \\
&
{ W}_{3,1} = \begin{pmatrix}   1 & 0 &0  \\ -\sqrt{2} \cot\frac{\theta}{2} & 0 & 0 \\ \cot^2\frac{\theta}{2} & 0 & 0\end{pmatrix} ,
{ W}_{3,2} = \begin{pmatrix}  0 & -\frac{\tan\frac{\theta}{2}}{\sqrt{2}} & 0 \\
0 & 1 &0 \\ 0 &  -\frac{\cot\frac{\theta}{2}}{\sqrt{2}} & 0\end{pmatrix} ,
{ W}_{3,3} = \begin{pmatrix}   0 & 0 & \tan^2\frac{\theta}{2} \\ 0 & 0 &  -\sqrt{2} \tan\frac{\theta}{2} \\
 0 & 0 & 1 \end{pmatrix} .
\end{eqnarray}
\end{widetext}
As in the case of spin one-half, some of the elements diverge at $\theta \to 0$.  Novel feature here is the divergence at $\theta \to \frac{\pi}{2}$ for the elements of $W_{2,2}$.  They are to be  compensated by the weight function to give finite answer to all physical quantities.
The weak values of the operator $L_z$ are given by
\begin{eqnarray}
(L_z)_{1,1} = 1, \quad (L_z)_{1,2} = 0,\quad (L_z)_{1,3} = -1,
\nonumber \\
(L_z)_{2,1} = 1, \quad (L_z)_{2,2} = 0,\quad (L_z)_{2,3} = -1,
\nonumber \\
(L_z)_{3,1} = 1, \quad (L_z)_{3,2} = 0,\quad (L_z)_{3,3} = -1,
\end{eqnarray}
which are easily understood because the pre-selection states $\psi_{j}$ are the eigenstates of this operator.
For $L_x$ we have the weak values
\begin{eqnarray}
&
(L_x)_{1,1} = \tan\frac{\theta}{2}, (L_x)_{1,2} = \frac{1}{\sin\theta},
(L_x)_{1,3} = \cot\frac{\theta}{2},
\nonumber \\
&
(L_x)_{2,1} = -\cot{\theta}, (L_x)_{2,2} = 0,
(L_x)_{2,3} = \cot{\theta},
\nonumber \\
&
(L_x)_{3,1} =  -\cot\frac{\theta}{2},  (L_x)_{3,2} = -\frac{1}{\sin\theta},
(L_x)_{3,3} = -\tan\frac{\theta}{2},
\nonumber \\
\end{eqnarray}
some of which diverge at $\theta \to 0$.
The weak values of rotated spin operator $L_\theta$ is naturally given by
\begin{eqnarray}
&
(L_\theta)_{1,1} = 1, \quad (L_\theta)_{1,2} = 1,\quad (L_\theta)_{1,3} = 1,
\nonumber \\
&
(L_\theta)_{2,1} = 0, \quad (L_\theta)_{2,2} = 0,\quad (L_\theta)_{2,3} = 0,
\nonumber \\
&
(L_\theta)_{3,1} = -1, \quad (L_\theta)_{3,2} = -1,\quad (L_\theta)_{3,3} = -1.
\end{eqnarray}
The divergence of $L_x$ at $\theta \to 0$ is understood from the relation
\begin{eqnarray}
L_x = - L_z \cot{\theta} +L_\theta {\rm \,cosec\,}{\theta} .
\end{eqnarray}
The weak values of $L_y$ are
\begin{eqnarray}
&
(L_y)_{1,1} = {\rm i} \tan\frac{\theta}{2},  (L_y)_{1,2} = -{\rm i} \cot\theta,
(L_y)_{1,3} = -{\rm i}\cot\frac{\theta}{2},
\nonumber \\
&
(L_y)_{2,1} = -{\rm i}\cot{\theta},  (L_y)_{2,2} = {\rm i}\tan\theta,
(L_y)_{2,3} = -{\rm i}\cot{\theta},
\nonumber \\
&
(L_y)_{3,1} =  -{\rm i}\cot\frac{\theta}{2},  (L_y)_{3,2} = -{\rm i}\cot{\theta},
(L_y)_{3,3} = {\rm i}\tan\frac{\theta}{2}.
 \nonumber \\
\end{eqnarray}
%
Thus, we find a very analogous structure of weak values to the one in the case of spin one-half: the divergence of  ``non-diagonal'' weak values of angular momentum operators at $\theta \to 0$.  The nontrivial divergence of $(L_y)_{2,2}$ at $\theta \to \frac{\pi}{2}$ is a novel feature of spin one system.  The fact that the behaviors found in $(L)_{\ell,j}$ is generic for all Hermitian operators in three-dimensional Hilbert space, can be checked by calculating the weak values of all Gell-Mann matrices,
%
%
with which (plus unit matrix) arbitrary Hermitian operators is uniquely decomposed.
%
%

\section{Reconstruction of quantum states before the measurement}

Suppose Alice passes on a mixed state
\begin{eqnarray}
\rho = \sum_j \left| \psi_j \right> \rho^{(\psi)}_j \left< \psi_j \right|
\end{eqnarray}
to Bob, on which Bob performs a projective measurement using the basis
$\{ \left| \phi_\ell \right>, \ell=1, 2,...,n \}$ and obtain the mixed state
\begin{eqnarray}
\tau = \sum_{\ell} \left| \phi_\ell \right> \tau^{(\phi)}_{\ell} \left< \phi_\ell \right| .
\end{eqnarray}
Let us ask how Bob can reconstruct the state $\rho$ with the knowledge that Alice had obtained
her state from a projective measurement in the  basis $\{ \left| \psi_j \right>, j=1, 2,...,n \}$.

If the Alice's state is expressed in the basis $\{ \left| \phi_\ell \right> \}$, a generic
representation with non-diagonal elements should be obtained, namely
\begin{eqnarray}
\rho = \sum_{\ell, m} \left| \phi_\ell \right> \rho^{(\phi)}_{\ell m} \left< \phi_m \right|
\end{eqnarray}
with $\rho^{(\phi)}_{\ell m} =  \left< \phi_\ell \right| \rho \left| \phi_m \right>$.
After the projective measurement, only diagonal components of this expression remain, and we should have
$\tau^{(\phi)}_{\ell} = \rho^{(\phi)}_{\ell \ell}$.

If we consider the {\it weak value of $\rho$} between states $\left| \phi_m \right>$ and $\left| \psi_j \right>$, then obtain
\begin{eqnarray}
\frac{ \left< \phi_m \right| \rho \left| \psi_j \right> } {  \left< \phi_m | \psi_j \right> } = \rho^{(\psi)}_j ,
\end{eqnarray}
for any $m$, since $\rho  \left| \psi_j \right> = \rho^{(\psi)}_j  \left| \psi_j \right>$, thus giving the formal answer to our question in terms of the weak values.
In order to obtain the explicit expression of $\rho^{(\psi)}_j$ in terms of $\tau^{(\phi)}_{m}$, we rewrite this equation, by inserting the complete set $\sum_j \left| \psi_j\right> \left< \psi_j\right|$ in front of $\rho$ in the LHS,
in the form
\begin{eqnarray}
\label{ee11}
\left< \phi_m | \psi_j \right> \rho^{(\psi)}_j - \sum_{\ell\ne m} \left< \phi_\ell | \psi_j \right> \rho^{(\phi)}_{m \ell}  = \left< \phi_m | \psi_j \right> \tau^{(\phi)}_{m} ,
\end{eqnarray}
which can be reformulated as $N^2$ linear equations indexed by $(m, j)$;
\begin{eqnarray}
\label{ee111}
A^{(m, j)}_{k,\ell} X_{k,\ell} = B^{(m, j)},
\end{eqnarray}
for $N^2$ unknown variables
\begin{eqnarray}
&
X_{k,\ell} = \rho^{(\psi)}_k \qquad  (k = \ell),
\nonumber \\
&
\qquad =  \rho^{(\phi)}_{k, \ell} \qquad  (k\ne \ell)
\end{eqnarray}
with
\begin{eqnarray}
&
A^{(m, j)}_{k,\ell} = \delta_{k,j}\delta_{\ell,j} \left< \phi_m | \psi_j \right> - \delta_{m,k}(1-\delta_{\ell, m}) \left< \phi_\ell | \psi_j \right>,
\nonumber \\
&
B^{(m, j)}=  \left< \phi_m | \psi_j \right> \tau^{(\phi)}_{m}.
\end{eqnarray}
Clearly, this gives the solution to the problem of state reconstruction.

We can further multiply $ \left< \psi_j | \phi_k \right>$ to the above equation from the right and sum up by $j$ to obtain
%
\begin{eqnarray}
\sum_j \left< \phi_m | \psi_j \right>\left< \psi_j | \phi_k \right> \rho^{(\psi)}_j  - (1-\delta_{m k}) \rho^{(\phi)}_{m k} = \tau^{(\phi)}_m \delta_{m k} ,
\nonumber \\
\end{eqnarray}
which splits into
\begin{eqnarray}
\label{reconstruction}
\sum_j \mu_{m j} \rho^{(\psi)}_j  = \tau^{(\phi)}_m  ,
\\
\sum_j \left< \phi_m | \psi_j \right>\left< \psi_j | \phi_k \right> \rho^{(\psi)}_j   =  \rho^{(\phi)}_{m k} ,
\end{eqnarray}
which are the explicit forms of linear equations that enable us to obtain $\rho^{(\psi)}_j$ and then $\rho^{(\phi)}_{m \ell}$ ($m\ne \ell$) from $\tau^{(\phi)}_\ell$ \cite{NC00}.

%
%

An illustrative example is in order.  Consider consecutive measurements of a spin one-half system by Alice and Bob.  We assume Alice's basis is given by
\begin{eqnarray}
\label{bspre}
\left| \psi_1 \right> = \begin{pmatrix}  1 \\ 0 \end{pmatrix},
\quad
\left| \psi_2 \right> = \begin{pmatrix}  0 \\ 1 \end{pmatrix},
\end{eqnarray}
and Bob's one by
\begin{eqnarray}
\label{bspost}
\left| \phi_1 \right> = \begin{pmatrix}  \cos\frac{\theta}{2} \\  \sin\frac{\theta}{2} \end{pmatrix},
\quad
\left| \phi_2 \right> = \begin{pmatrix}   -\sin\frac{\theta}{2}  \\  \cos\frac{\theta}{2}  \end{pmatrix},
\end{eqnarray}
with the parameter range $0  \leqslant \theta\leqslant \pi$.
Alice's state
\begin{eqnarray}
&
\rho = \left| \psi_1 \right> \rho^{(\psi)}_1 \left< \psi_1 \right| + \left| \psi_2 \right> \rho^{(\psi)}_2 \left< \psi_2 \right|
\qquad\qquad\qquad\quad
\nonumber \\
&
\quad =
\left| \phi_1 \right> \tau^{(\phi)}_1 \left< \phi_1 \right| + \left| \phi_2 \right> \tau^{(\phi)}_2 \left< \phi_2 \right|
\qquad\qquad\qquad\qquad\nonumber \\
&
\qquad\qquad
+ \left| \phi_1 \right> \rho^{(\phi)}_{12} \left< \phi_2 \right|
+ \left| \phi_2 \right> \rho^{(\phi)}_{21} \left< \phi_1 \right|
\end{eqnarray}
is projected to
\begin{eqnarray}
\tau = \left| \phi_1 \right> \tau^{\phi}_1 \left< \phi_1 \right| + \left| \phi_2 \right> \tau^{\phi}_2 \left< \phi_2 \right|
\end{eqnarray}
by Bob's measurement.
Suppose that Bob wants to reconstruct the Alice's state $\rho$ from his state $\tau$ with the knowledge of the basis sets $\{\phi_1, \phi_2\}$ and $\{ \psi_1, \psi_2\}$.
Denoting a matrix $G_{\ell j} = \left< \phi_\ell | \psi_j \right>$ we have
\begin{eqnarray}
G = \begin{pmatrix} \cos\frac{\theta}{2} & \sin\frac{\theta}{2} \\ -\sin\frac{\theta}{2} & \cos\frac{\theta}{2}  \end{pmatrix}.
\label{2dg}
\end{eqnarray}
The equation (\ref{ee11}) takes the form
\begin{eqnarray}
\!\!\!\!
  \begin{pmatrix} G_{11} & 0 & -G_{21} & 0 \\ 0 & G_{12} & -G_{22} & 0 \\
  G_{21} & 0 & 0 & -G_{11} \\ 0 & G_{22} & 0 & -G_{12}  \end{pmatrix}
 \begin{pmatrix} \rho^{(\psi)}_1 \\ \rho^{(\psi)}_2 \\ \rho^{(\phi)}_{12} \\ \rho^{(\phi)}_{21}  \end{pmatrix}
  =  \begin{pmatrix} G_{11} \tau^{\phi}_1 \\ G_{12} \tau^{\phi}_1 \\ G_{21} \tau^{\phi}_2 \\ G_{22} \tau^{\phi}_2 \end{pmatrix},
\end{eqnarray}
whose solution is given by
\begin{eqnarray}
\begin{pmatrix} \rho^{(\psi)}_1 \\ \rho^{(\psi)}_2 \\ \rho^{(\phi)}_{12} \\ \rho^{(\phi)}_{21} \end{pmatrix}
 =
 \begin{pmatrix} \frac{\tau^{\phi}_{1}+\tau^{\phi}_{2}}{2} + \frac{\tau^{\phi}_{1}-\tau^{\phi}_{2}}{2} {\rm sec\,} \theta
  \\ \frac{\tau^{\phi}_{1}+\tau^{\phi}_{2}}{2} - \frac{\tau^{\phi}_{1}-\tau^{\phi}_{2}}{2} {\rm sec\,} \theta \\
 - \frac{\tau^{\phi}_{1}-\tau^{\phi}_{2}}{2} \tan \theta \\  - \frac{\tau^{\phi}_{1}-\tau^{\phi}_{2}}{2} \tan \theta \end{pmatrix} .
\end{eqnarray}
%
This expression is singular at the value of the angle parameter $\theta=\frac{\pi}{2}$ , signifying the
{\it irreversibility of the projective measurement}.   It means that the
information on the past history is completely erased with successive projective measurements with bases (\ref{bspre}) and (\ref{bspost}) with $\theta=\frac{\pi}{2}$.  We consider the generalization of this results to the system with higher dimensional Hilbert space.
 \section{Degenerate matrices of Birkhoff's polytope }
 The reconstruction of Alice's state by the results of Bob's measurement in the case of arbitrary spin can be performed, in principle, in the same manner with eq. (\ref{reconstruction}), but in reality, the task is nontrivial.
We need to characterize all permissible matrices with positive matrix elements $\mu_{\ell,j}$, which make valid our computations and allow us to solve the system of equations (\ref{reconstruction}).

First of all we have a condition
\begin{eqnarray}
\sum_{\ell}\mu_{\ell,j} =
\sum_{j}\mu_{\ell,j}=1.
\end{eqnarray}
The matrices, which are satisfied to such conditions, are called bistochastic or doubly stochastic.
The class of  $N \times N$ bistochastic matrices is a $(N-1)^2$ dimensional compact convex polyhedron known as the
Birkhoff's polytope $\mathcal{B}_N$ \cite{Bi46, Brualdi}. The distance between two matrices is defined by
\begin{equation}
D(A,B)=\sqrt{\mathrm{Tr}(A-B)(A^\dag-B^\dag)}.
\end{equation}
The boundary consists of corners, edges, faces, 3-faces and so on.
The extreme points or corners of the polytope represent permutation matrices $P^{(N)}$.

At first let us summarize some well-known properties of two-dimensional and three-dimensional bistochastic matrices.

In the case of $N=2$, $\mathcal{B}_2$ is a line segment with the endpoints corresponding to permutation matrices
\begin{eqnarray}
P^{(2)}_{0}=\begin{pmatrix} 1 & 0 \\ 0 & 1 \end{pmatrix},
\quad
P^{(2)}_{1}=\begin{pmatrix} 0 & 1 \\ 1 & 0 \end{pmatrix}.
\end{eqnarray}
The distance between these endpoints is equal to $D(P^{(2)}_{0},P^{(2)}_{1})=2$.
Any bisochastic matrix $\mu^{(2)}$ inside that line can be formed by combination
\begin{eqnarray}
\mu^{(2)}=p_0 P^{(2)}_{0}+p_1 P^{(2)}_{1},
\end{eqnarray}
with conditions
\begin{eqnarray}
p_0+p_1=1,\;\;\;\;\;\; 0\leq p_0 \leq 1,\;\;\;\;\;\; 0\leq p_1 \leq 1.
\end{eqnarray}
If we use a parametrization $p_0=\cos^2 \frac{\theta}{2}$, $p_1=\sin^2 \frac{\theta}{2}$, where $0\leq \theta \leq \pi$,
then obtain
\begin{eqnarray}
\mu^{(2)}=\begin{pmatrix} \cos^2 \frac{\theta}{2} & \sin^2 \frac{\theta}{2} \\ \sin^2 \frac{\theta}{2} &  \cos^2 \frac{\theta}{2}\end{pmatrix}.
\label{mutheta}
\end{eqnarray}

In the case of $N=3$ the Birkhoff's polytope contains 6 corners of permutation matrices

\begin{eqnarray}
P^{(3)}_0=\begin{pmatrix} 1 & 0 & 0 \\ 0 & 1 & 0 \\ 0 & 0 & 1 \end{pmatrix},
P^{(3)}_1=\begin{pmatrix} 1 & 0 & 0 \\ 0 & 0 & 1 \\ 0 & 1 & 0 \end{pmatrix},
P^{(3)}_2=\begin{pmatrix} 0 & 1 & 0 \\ 1 & 0 & 0 \\ 0 & 0 & 1 \end{pmatrix},
\nonumber \\
P^{(3)}_3=\begin{pmatrix} 0 & 1 & 0 \\ 0 & 0 & 1 \\ 1 & 0 & 0 \end{pmatrix},
P^{(3)}_4=\begin{pmatrix} 0 & 0 & 1 \\ 1 & 0 & 0 \\ 0 & 1 & 0 \end{pmatrix},
P^{(3)}_5=\begin{pmatrix} 0 & 0 & 1 \\ 0 & 1 & 0 \\ 1 & 0 & 0 \end{pmatrix}.
\nonumber \\
\end{eqnarray}
It is easy to check that there are 6 longer edges with lengths
$D(P^{(3)}_0,P^{(3)}_3)=D(P^{(3)}_0,P^{(3)}_4)=D(P^{(3)}_3,P^{(3)}_4)=
D(P^{(3)}_1,P^{(3)}_2)=D(P^{(3)}_1,P^{(3)}_5)=D(P^{(3)}_2,P^{(3)}_5)=\sqrt{6}$,
which form two equilateral triangles placed in two orthogonal 2-planes.
The other 9 edges are shorter and have a length 2.
An arbitrary bistochastic matrix inside $\mathcal{B}_3$ can be represented by

\begin{eqnarray}
\mu^{(3)}=\sum_{i=0}^{5}p_i P^{(3)}_i,
\label{mu3}
\end{eqnarray}
with condition
\begin{eqnarray}
\sum_{i=0}^{5} p_i=1\;\;\;\;\; (0\leq p_i \leq 1).
\label{cond3}
\end{eqnarray}
For any matrix $\mu^{(3)}$ the representation (\ref{mu3}) is not unique
as the dimension of the space of $3 \times 3$ bistochastic matrices is $4$ and we have 6
parameters connected with the condition (\ref{cond3}). Though the different points inside
Birkhoff's polytope can correspond to the same bistochastic matrix, the representation
(\ref{mu3}) is convenient and useful to characterize a space of bistochastic matrices.

\begin{figure}[tbp]
\includegraphics[width=7cm]{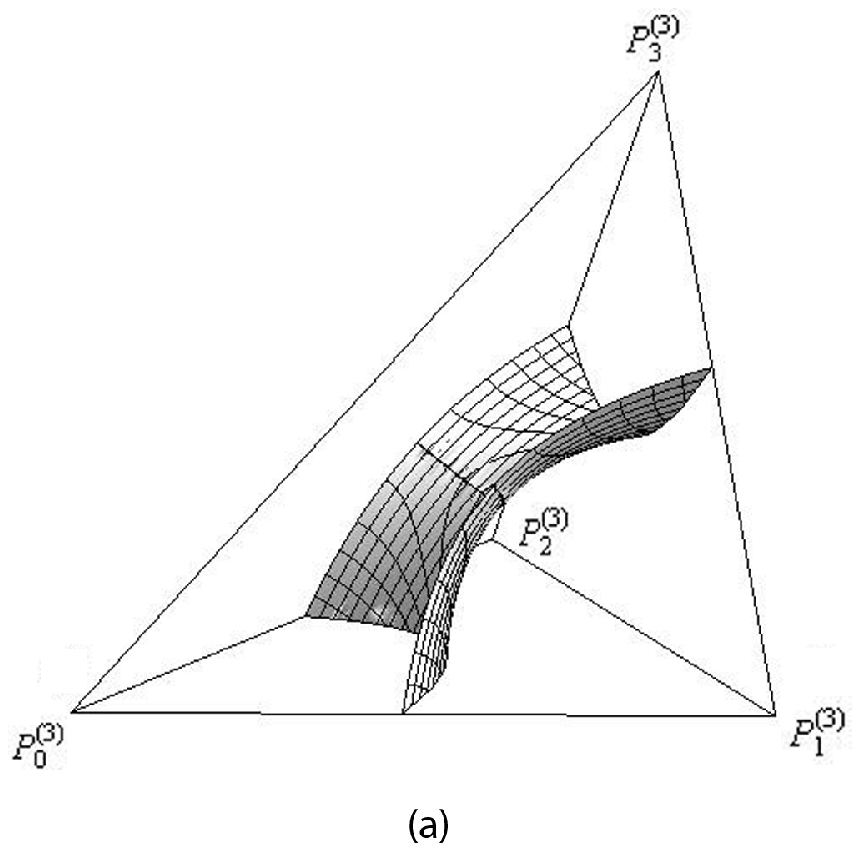}
\includegraphics[width=7cm]{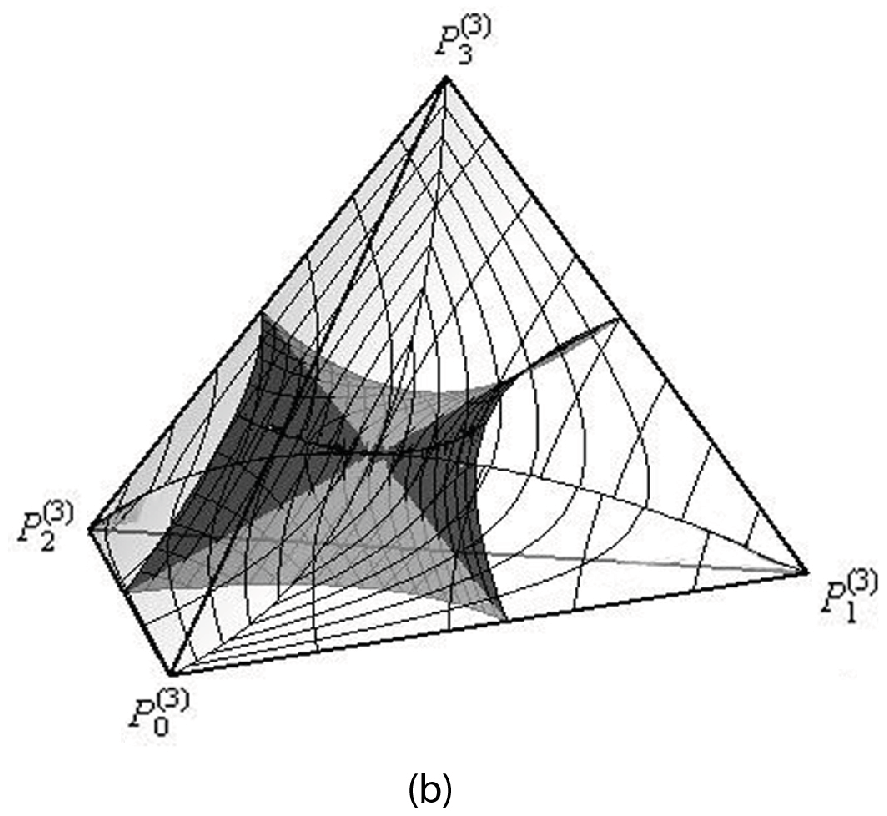}
\centering
\caption{
The surface of degenerate matrices of the 3-plain $P_0^{(3)}P_1^{(3)}P_2^{(3)}P_3^{(3)}$
of the Bikhoff's polytope (a) and it's intersection with the unistochastic surface (b).
The intersection consists of two lines $O_1 O'_1$ and $O_2 O'_2$, connecting the centers of edges
of irregular tetrahedron.
}
\label{fig1}
\end{figure}

The second condition for reconstructing an initial information after quantum measurement
is the existence of unitary matrix $G$,
which constrict our set, but does not decrease a dimension. Bistochastic matrices,
which can be represented by $\{\mu_{\ell,j} = \left| {\left< \phi_\ell | \psi_j \right>} \right| ^2 \}$ are
called unistochastic. In general case for arbitrary $N$ there is no certain way to
check whether the given bistochastic matrix is unistochastic or not. However, for $N=2$
the answer is obvious, since the all $2 \times 2$ bistochastic matrices (\ref{mutheta}) are unistochastic
and the corresponding $G$ matrix can be chosen as (\ref{2dg}). For the case $N=3$ it is
always possible to check whether the given bistochastic matrix is unistochastic or not.
Introducing new notations
\begin{eqnarray}
&
L_1=\sqrt{\mu_{11}^{(3)}\mu_{12}^{(3)}},\
L_2=\sqrt{\mu_{21}^{(3)}\mu_{22}^{(3)}},
\nonumber \\
&
L_3=\sqrt{\mu_{31}^{(3)}\mu_{32}^{(3)}},
\end{eqnarray}
we verify a condition of forming triangle with side lengths $L_1,L_2$ and $L_3$
\begin{eqnarray}
\left|L_2-L_3\right|\leq L_1\leq L_2 + L_3.
\label{conduni}
\end{eqnarray}
If the inequalities (\ref{conduni}) are satisfied, then the matrix $\mu^{(3)}$ is unistochastic.
The unistochastic subset $\mathcal{U}_3$ of $\mathcal{B}_3$ was studied in \cite{Bengtsson}.

\begin{figure}[tbp]
\includegraphics[width=7cm]{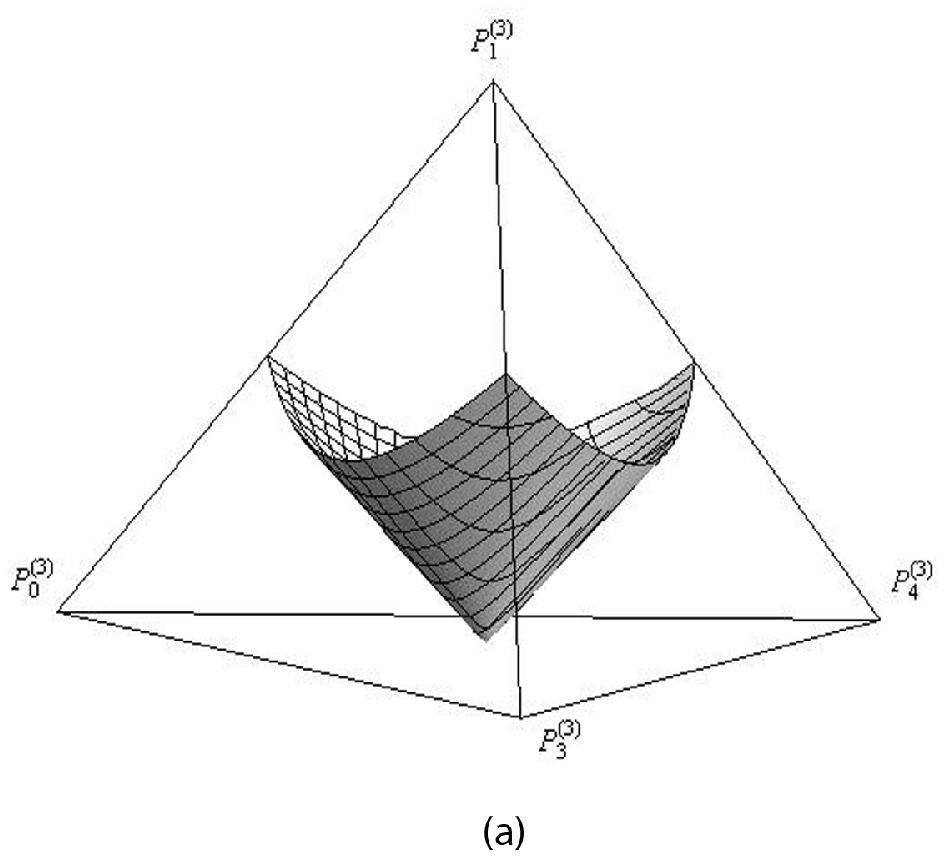}
\includegraphics[width=7cm]{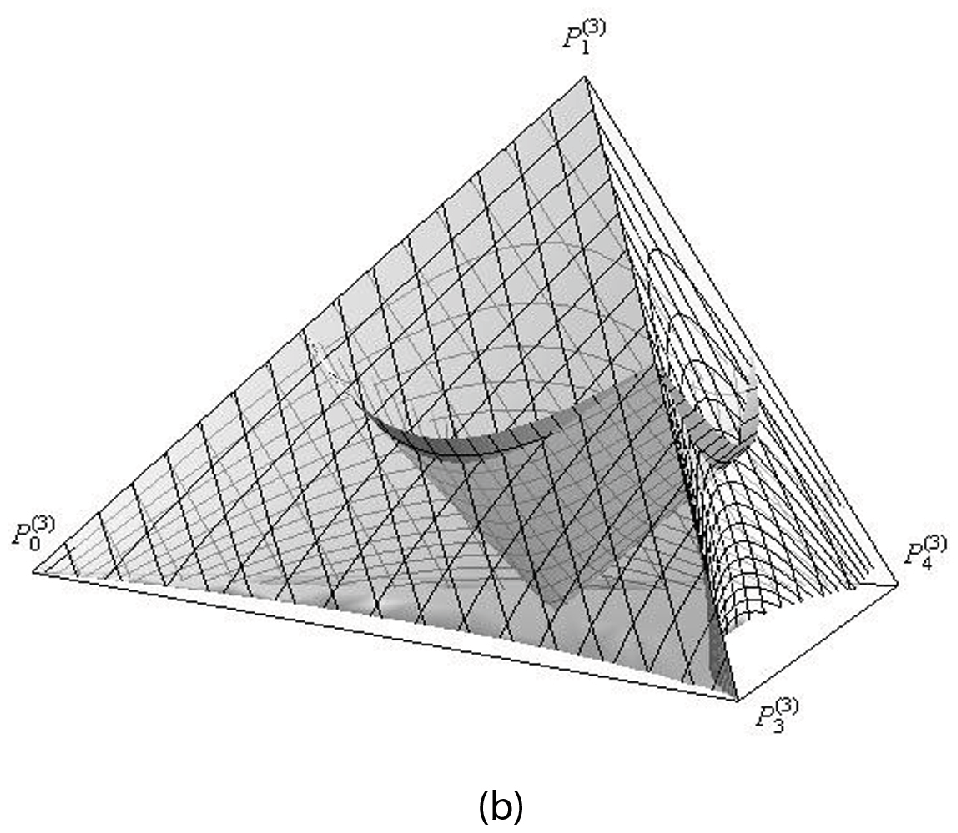}
\centering
\caption{The surface of degenerate matrices of the facet $P_0^{(3)}P_1^{(3)}P_3^{(3)}P_4^{(3)}$ (a)
and it's intersection with the 3-surface of unistochastic matrices (b).
The center of equilateral triangle $P_1^{(3)}P_3^{(3)}P_4^{(3)}$ belongs to the surface degenerate matrices.
The boundary of unistocastic matrices on the plain $P_1^{(3)}P_3^{(3)}P_4^{(3)}$ is a 3-hypocycloid.
}
\label{fig2}
\end{figure}

The third condition for obtaining coefficients $P_j$ is the existence of unique solution of the system
of linear equations (\ref{reconstruction}), which means that the matrix $\mu$ has to be invertible.
For the case $N=2$ the matrix (\ref{mutheta}) is degenerate only if $\theta=\pi/2$, which corresponds
to the midpoint of the segment of bistochastic matrices.   This amounts to the reconfirmation of the argument of irreversibility in the previous section.
When $N=3$ we have a three-dimensional surface
of degenerate bistochastic matrices, which is specified by the condition
\begin{eqnarray}
\det  \mu^{(3)} = 0 .
\end{eqnarray}
Notice that the center of Birkhoff's polytope $\mu^{(3)}_{ij}=\frac{1}{3}$
also belongs to the surface of degeneracy. To characterize this surface we depict its boundaries in
corresponding three-dimensional surfaces of Birkhoff's polytope. Figure (\ref{fig1}a) illustrates
a surface of degenerate bistochastic matrices, which have components $p_4=0$ and $p_5=0$ in the
representation (\ref{mu3}). Figure (\ref{fig1}b) demonstrates an intersection of the surfaces of degenerate
and unistochastic matrices. This intersection consists of two lines $O_1 O_1'$ and $O_2 O_2'$, where $O_1,O_1',O_2,O_2'$
are the midpoints of the edges $P_0^{(3)}P_2^{(3)},P_1^{(3)}P_3^{(3)},P_2^{(3)}P_3^{(3)},P_0^{(3)}P_1^{(3)}$ correspondingly.
Thus, to obtain a set of permissible matrices $\mu^{(3)}$, we have to subtract the lines $O_1 O_1'$ and $O_2 O_2'$
from the surface of unistochastic matrices. In figure (\ref{fig2}a) the surface of non-invertible matrices is shown
within the 3-plain $P_0^{(3)}P_1^{(3)}P_3^{(3)}P_4^{(3)}$. It touches a plain $P_0^{(3)}P_3^{(3)}P_4^{(3)}$  at
the center of the equilateral triangle. Note that the centers of segments
$P_0^{(3)}P_1^{(3)}$,$\;P_3^{(3)}P_1^{(3)}$,$\;P_4^{(3)}P_1^{(3)}$ also belong to this surface. The set of
unistochastic matrices with components $p_2=0$ and $p_5=0$ represent a three-dimensional volume,
which contains edges $P_0^{(3)}P_1^{(3)}$,$\;P_3^{(3)}P_1^{(3)}$,$\;P_4^{(3)}P_1^{(3)}$. On the plain
$P_0^{(3)}P_3^{(3)}P_4^{(3)}$ the boundary of unistochastic subset is the famous hypocycloid \cite{hypo}.
The intersection of the sets of unistochastic and degenerate matrices is shown in figure (\ref{fig2}b).
Here also the set of permissible matrices $\mu^{(3)}$ can be obtained by subtracting the set of non-invertible
matrices from the volume of unistochastic ones.

\section{Summary and Prospects}

In this article we have shown that the weak values emerge quite naturally from the gauge invariant expansion of Hermitian operators using two sets of orthonormal bases.  The absence of the smooth single orthonormal basis in the limit $ \{ \phi_\ell \} \to \{ \psi_\ell \}$ of the expansion seems to explain the reason why the concept of the weak value has eluded the discovery by all practitioners of quantum mechanics until late 1980s.

It will be both very interesting mathematically and useful experimentally to characterize the unistochastic matrices of higher dimension and their irreversible subsets within the Birkhoff's polytope.  It appears, however, that we have no general recipe for this task at this point, since characterizing the structure of the Birkhoff's polytope itself is already a hard task, partially completed only up to $N=4$ \cite{Bengtsson}.

\section*{Acknowledgments}
We thank Prof. I. Tsutsui, Prof. A. Tanaka, Dr. N. Yonezawa, and Prof. Y. Shikano for stimulating discussions. This research was supported  by the Japan Ministry of Education, Culture, Sports, Science and Technology under the Grant number 24540412.



\begin{thebibliography}{99}

\bibitem{AA88}
 Y. Aharonov, D. Z. Albert and L. Vaidman,
 Phys. Rev. Lett. {\bf 60}, 1351 (1988).

\bibitem{AA90}
Y. Aharonov and L. Vaidman, Phys. Rev. A {\bf 41}, 11 (1990).

\bibitem{RHS91}
N. W. M. Ritchie, J. G. Story, and R. G. Hulet, Phys. Rev. Lett. 66, 1107 (1991).

\bibitem{RA95}
B. Reznik and Y. Aharonov, Phys. Rev. A {\bf 52}, 2538 (1995).

\bibitem{AA2005}
Y. Aharonov and A. Botero, Phys. Rev. A {\bf 72}, 052111 (2005).

\bibitem{Aharonov_1964}
Y. Aharonov, P. G. Bergmann and L. Lebowitz,
\newblock Phys. Rev. {\bf 134}, B1410 (1964).

\bibitem{AP09}
Y. Aharonov, S. Popescu, J. Tollaksen and L. Vaidman,
Phys. Rev. A {\bf 79}, 052110 (2009).

\bibitem{BR2000}
A. Botero and B. Reznik, Phys. Rev. A {\bf 61}, 050301(R) (2000).

\bibitem{ABPR2002}
Y. Aharonov, A. Botero, S. Popescu and B. Reznik, Phys. Lett. A {\bf 301}, 130 (2002).

\bibitem{HLS11}
E. Haapasalo, P. Lahti and J. Schultz, Phys. Rev. A {\bf 84}, 052107 (2011).

\bibitem{FF12}
J. Fischbach and M. Freyberger, Phys. Rev. A {\bf 86}, 052110 (2012).

\bibitem{Hof12}
H. F. Hofmann, Phys. Rev. Lett. {\bf 109}, 020408 (2012).

\bibitem{Ta02}
A. Tanaka, Phys. Lett. A {\bf 297}, 307 (2002).

\bibitem{SH10}
 Y. Shikano and A. Hosoya, J. Phys. A: Math. Theor.  {\bf 43}, 025304 (2010). 

\bibitem{MS12}
 T. Morita, T. Sasaki and I. Tsutsui, Prog. Theor. Exp. Phys. {\bf 2013}, 053A02 (2013).

\bibitem{Bi46}
G. Birkhoff, Univ. Nac. Tucum\'{a}n Rev. A {\bf 5}, 147 (1946).

%














%

\bibitem{NC00}
M. A. Nielsenand I. L. Chuang, {\it Quantum computation and quantum information}, (Cambridge U.P., 2000).

\bibitem{Brualdi}
R. A. Brualdi and P. M. Gibson, J. Comb. Theory A {\bf 22}, 194 (1977).

\bibitem{Bengtsson}
I. Bengtson, {\AA}. Ericsson, M. Ku\'{s}, W. Tadej and K. \.{Z}yczkowski,
Commun. Math. Phys. {\bf 259}, 307 (2005).

\bibitem{hypo}
K. \.{Z}yczkowski, M. Ku\'{s}, W. S{\l}omczynski and H.-J. Sommers,
J. Phys. A {\bf 36}, 3425 (2003).

\end{thebibliography}
\end{document}